\documentclass[aps,prl,twocolumn,showpacs,groupedaddress]{revtex4}  
\usepackage{graphicx}  
\usepackage{dcolumn}   
\usepackage{bm}        
\usepackage{amssymb}   

\begin{document}
\newcommand{\met}{\not{\!\!\!E_T}}
\newcommand{\ttbar}     {\mbox{$t\bar{t}$}}
\newcommand{\bbbar}     {\mbox{$b\bar{b}$}}
\newcommand{\ccbar}     {\mbox{$c\bar{c}$}}
\newcommand{\ptrel}     {\mbox{$p^{\text{rel}}_T$}}
\newcommand{\fq}        {\rho}

\leftline{\hspace{0.2in}\mbox{PRL. {\bf98}, 041801 (2007)} \hspace{3.4in} \mbox{FERMILAB-PUB-06/278-E} }

\title{Experimental discrimination between charge $2e/3$ top quark and charge $4e/3$ exotic quark production scenarios}

%
\author{                                                                      
V.M.~Abazov,$^{36}$                                                           
B.~Abbott,$^{76}$                                                             
M.~Abolins,$^{66}$                                                            
B.S.~Acharya,$^{29}$                                                          
M.~Adams,$^{52}$                                                              
T.~Adams,$^{50}$                                                              
M.~Agelou,$^{18}$                                                             
S.H.~Ahn,$^{31}$                                                              
M.~Ahsan,$^{60}$                                                              
G.D.~Alexeev,$^{36}$
% ...
\\                                                                            
\vskip 0.30cm                                                                 
\centerline{(D\O\ Collaboration)}                                             
\vskip 0.30cm                                                                 
}                                                                             
\affiliation{                                                                 
\centerline{$^{1}$Universidad de Buenos Aires, Buenos Aires, Argentina}       
\centerline{$^{2}$LAFEX, Centro Brasileiro de Pesquisas F{\'\i}sicas,         
                  Rio de Janeiro, Brazil}                                     
\centerline{$^{3}$Universidade do Estado do Rio de Janeiro,                   
                  Rio de Janeiro, Brazil}                                     
\centerline{$^{4}$Instituto de F\'{\i}sica Te\'orica, Universidade            
                  Estadual Paulista, S\~ao Paulo, Brazil}                     
\centerline{$^{5}$University of Alberta, Edmonton, Alberta, Canada,           
                  Simon Fraser University, Burnaby, British Columbia, Canada,}
\centerline{York University, Toronto, Ontario, Canada, and                    
                  McGill University, Montreal, Quebec, Canada}                
\centerline{$^{6}$Institute of High Energy Physics, Beijing,                  
                  People's Republic of China}                                 
\centerline{$^{7}$University of Science and Technology of China, Hefei,       
                  People's Republic of China}                                 
\centerline{$^{8}$Universidad de los Andes, Bogot\'{a}, Colombia}             
\centerline{$^{9}$Center for Particle Physics, Charles University,            
                  Prague, Czech Republic}                                     
\centerline{$^{10}$Czech Technical University, Prague, Czech Republic}        
\centerline{$^{11}$Center for Particle Physics, Institute of Physics,         
                   Academy of Sciences of the Czech Republic,                 
                   Prague, Czech Republic}                                    
\centerline{$^{12}$Universidad San Francisco de Quito, Quito, Ecuador}        
\centerline{$^{13}$Laboratoire de Physique Corpusculaire, IN2P3-CNRS,         
                   Universit\'e Blaise Pascal, Clermont-Ferrand, France}      
\centerline{$^{14}$Laboratoire de Physique Subatomique et de Cosmologie,      
                   IN2P3-CNRS, Universite de Grenoble 1, Grenoble, France}    
\centerline{$^{15}$CPPM, IN2P3-CNRS, Universit\'e de la M\'editerran\'ee,     
                   Marseille, France}                                         
\centerline{$^{16}$IN2P3-CNRS, Laboratoire de l'Acc\'el\'erateur              
                   Lin\'eaire, Orsay, France}                                 
\centerline{$^{17}$LPNHE, IN2P3-CNRS, Universit\'es Paris VI and VII,         
                   Paris, France}                                             
\centerline{$^{18}$DAPNIA/Service de Physique des Particules, CEA, Saclay,    
                   France}                                                    
\centerline{$^{19}$IPHC, IN2P3-CNRS, Universit\'e Louis Pasteur, Strasbourg,  
                    France, and Universit\'e de Haute Alsace,                 
                    Mulhouse, France}                                         
\centerline{$^{20}$Institut de Physique Nucl\'eaire de Lyon, IN2P3-CNRS,      
                   Universit\'e Claude Bernard, Villeurbanne, France}         
\centerline{$^{21}$III. Physikalisches Institut A, RWTH Aachen,               
                   Aachen, Germany}                                           
\centerline{$^{22}$Physikalisches Institut, Universit{\"a}t Bonn,             
                   Bonn, Germany}                                             
\centerline{$^{23}$Physikalisches Institut, Universit{\"a}t Freiburg,         
                   Freiburg, Germany}                                         
\centerline{$^{24}$Institut f{\"u}r Physik, Universit{\"a}t Mainz,            
                   Mainz, Germany}                                            
\centerline{$^{25}$Ludwig-Maximilians-Universit{\"a}t M{\"u}nchen,            
                   M{\"u}nchen, Germany}                                      
\centerline{$^{26}$Fachbereich Physik, University of Wuppertal,               
                   Wuppertal, Germany}                                        
\centerline{$^{27}$Panjab University, Chandigarh, India}                      
\centerline{$^{28}$Delhi University, Delhi, India}                            
\centerline{$^{29}$Tata Institute of Fundamental Research, Mumbai, India}     
\centerline{$^{30}$University College Dublin, Dublin, Ireland}                
\centerline{$^{31}$Korea Detector Laboratory, Korea University,               
                   Seoul, Korea}                                              
\centerline{$^{32}$SungKyunKwan University, Suwon, Korea}                     
\centerline{$^{33}$CINVESTAV, Mexico City, Mexico}                            
\centerline{$^{34}$FOM-Institute NIKHEF and University of                     
                   Amsterdam/NIKHEF, Amsterdam, The Netherlands}              
\centerline{$^{35}$Radboud University Nijmegen/NIKHEF, Nijmegen, The          
                  Netherlands}                                                
\centerline{$^{36}$Joint Institute for Nuclear Research, Dubna, Russia}       
\centerline{$^{37}$Institute for Theoretical and Experimental Physics,        
                   Moscow, Russia}                                            
\centerline{$^{38}$Moscow State University, Moscow, Russia}                   
\centerline{$^{39}$Institute for High Energy Physics, Protvino, Russia}       
\centerline{$^{40}$Petersburg Nuclear Physics Institute,                      
                   St. Petersburg, Russia}                                    
\centerline{$^{41}$Lund University, Lund, Sweden, Royal Institute of          
                   Technology and Stockholm University, Stockholm,            
                   Sweden, and}                                               
\centerline{Uppsala University, Uppsala, Sweden}                              
\centerline{$^{42}$Physik Institut der Universit{\"a}t Z{\"u}rich,            
                   Z{\"u}rich, Switzerland}                                   
\centerline{$^{43}$Lancaster University, Lancaster, United Kingdom}           
\centerline{$^{44}$Imperial College, London, United Kingdom}                  
\centerline{$^{45}$University of Manchester, Manchester, United Kingdom}      
\centerline{$^{46}$University of Arizona, Tucson, Arizona 85721, USA}         
\centerline{$^{47}$Lawrence Berkeley National Laboratory and University of    
                   California, Berkeley, California 94720, USA}               
\centerline{$^{48}$California State University, Fresno, California 93740, USA}
\centerline{$^{49}$University of California, Riverside, California 92521, USA}
\centerline{$^{50}$Florida State University, Tallahassee, Florida 32306, USA} 
\centerline{$^{51}$Fermi National Accelerator Laboratory,                     
            Batavia, Illinois 60510, USA}                                     
\centerline{$^{52}$University of Illinois at Chicago,                         
            Chicago, Illinois 60607, USA}                                     
\centerline{$^{53}$Northern Illinois University, DeKalb, Illinois 60115, USA} 
\centerline{$^{54}$Northwestern University, Evanston, Illinois 60208, USA}    
\centerline{$^{55}$Indiana University, Bloomington, Indiana 47405, USA}       
\centerline{$^{56}$University of Notre Dame, Notre Dame, Indiana 46556, USA}  
\centerline{$^{57}$Purdue University Calumet, Hammond, Indiana 46323, USA}    
\centerline{$^{58}$Iowa State University, Ames, Iowa 50011, USA}              
\centerline{$^{59}$University of Kansas, Lawrence, Kansas 66045, USA}         
\centerline{$^{60}$Kansas State University, Manhattan, Kansas 66506, USA}     
\centerline{$^{61}$Louisiana Tech University, Ruston, Louisiana 71272, USA}   
\centerline{$^{62}$University of Maryland, College Park, Maryland 20742, USA} 
\centerline{$^{63}$Boston University, Boston, Massachusetts 02215, USA}       
\centerline{$^{64}$Northeastern University, Boston, Massachusetts 02115, USA} 
\centerline{$^{65}$University of Michigan, Ann Arbor, Michigan 48109, USA}    
\centerline{$^{66}$Michigan State University,                                 
            East Lansing, Michigan 48824, USA}                                
\centerline{$^{67}$University of Mississippi,                                 
            University, Mississippi 38677, USA}                               
\centerline{$^{68}$University of Nebraska, Lincoln, Nebraska 68588, USA}      
\centerline{$^{69}$Princeton University, Princeton, New Jersey 08544, USA}    
\centerline{$^{70}$State University of New York, Buffalo, New York 14260, USA}
\centerline{$^{71}$Columbia University, New York, New York 10027, USA}        
\centerline{$^{72}$University of Rochester, Rochester, New York 14627, USA}   
\centerline{$^{73}$State University of New York,                              
            Stony Brook, New York 11794, USA}                                 
\centerline{$^{74}$Brookhaven National Laboratory, Upton, New York 11973, USA}
\centerline{$^{75}$Langston University, Langston, Oklahoma 73050, USA}        
\centerline{$^{76}$University of Oklahoma, Norman, Oklahoma 73019, USA}       
\centerline{$^{77}$Oklahoma State University, Stillwater, Oklahoma 74078, USA}
\centerline{$^{78}$Brown University, Providence, Rhode Island 02912, USA}     
\centerline{$^{79}$University of Texas, Arlington, Texas 76019, USA}          
\centerline{$^{80}$Southern Methodist University, Dallas, Texas 75275, USA}   
\centerline{$^{81}$Rice University, Houston, Texas 77005, USA}                
\centerline{$^{82}$University of Virginia, Charlottesville,                   
            Virginia 22901, USA}                                              
\centerline{$^{83}$University of Washington, Seattle, Washington 98195, USA}  
}                                                                             

\date{August 16, 2006}

\begin{abstract}
We present the first experimental discrimination 
between the $2e/3$ and $4e/3$ top quark electric charge scenarios, 
using top quark pairs ($\ttbar$) produced in $p\bar{p}$
collisions at $\sqrt{s}$=1.96 TeV by the Fermilab Tevatron collider.
We use 370 pb$^{-1}$ of data collected by the D0 experiment and
select events with at least one high transverse momentum electron
or muon, high transverse energy imbalance, and four or more jets.
We discriminate between $b$- and $\bar{b}$-quark jets by using
the charge and momenta of tracks within the jet cones.
The data is consistent with the expected electric charge,
$|q|=2e/3$. We exclude, at the 92\% C.L., that the sample is
solely due to the production of exotic quark pairs $Q\bar{Q}$ with
$|q|=4e/3$. We place an upper limit on the fraction of $Q\bar{Q}$
pairs $\fq < 0.80$ at the 90\% C.L.
\end{abstract}

\maketitle

The heavy particle discovered by the CDF and D0 collaborations 
at the Fermilab Tevatron proton-antiproton collider in 
1995~\cite{discovery} is widely recognized to be the top quark. 
Currently measured properties of the particle are consistent with standard
model (SM) expectations for the top quark. However, many of the
properties of the particle are still poorly known. In particular,
its electric charge, a fundamental quantity characterizing a
particle, has not yet been determined. 

To date, it is possible to interpret the discovered particle as either 
a charge $2e/3$ or $-4e/3$ quark. In the published top quark analyses 
of the CDF and D0 collaborations~\cite{top_review}, there is a 
two-fold ambiguity in pairing the $b$-quarks and the $W$ bosons in 
the reaction  $p \bar{p} \to t \bar{t}\to W^+W^- b \bar{b}$, 
and equivalently, in the electric charge assignment of the measured 
particle. In addition to the SM assignment, $t \to W^+ b$, ``$t$''$ \to
W^-b$ is also conceivable, in which case ``$t$'' would
actually be an exotic quark, $Q$, with charge $q=-4e/3$ (charge-conjugate 
processes are implied). 
It is possible to fit $Z\to \ell^+ \ell^-$ and $Z\to
b \bar{b}$ data assuming a top quark mass of $m_{t}=270$~GeV and 
a right-handed $b$-quark that mixes with the isospin +1/2
component of an exotic doublet of charge $-1e$/3 and $-4 e$/3
quarks, $(Q1~,Q4)_{R}$~\cite{exotic_top_paper}. In this scenario,
the $-4 e$/3 charge quark is the particle discovered at the 
Tevatron, and the top quark, with mass of 270~GeV, would have so far
escaped detection.

In this Letter, we report the first experimental discrimination
between the $2e/3$ and $4e/3$ charge scenarios. We also consider the case 
where the analyzed sample contains an admixture of SM top quarks 
and exotic quarks and place an upper limit on the exotic quark fraction. 
Our search strategy assumes each quark decays 100\% of the time to a $W$ boson 
and a $b$-quark. We use the lepton-plus-jets channel which arises 
when one $W$ boson decays leptonically and one decays hadronically. 
The charged leptons ($e/\mu$) originate from a direct $W$ decay or
from $W \rightarrow \tau \rightarrow e/\mu$. We require that the 
final state have at least two $b$-quark jets. The data used in this 
Letter were collected by the D0 experiment from June 2002 through 
August 2004 and correspond to an integrated luminosity of 370 pb$^{-1}$.


The D0 detector includes a tracking system, calorimeters, and a
muon spectrometer~\cite{d0det}. The tracking system is made up of a
silicon microstrip tracker (SMT) and a central fiber tracker,
located inside a 2~T superconducting solenoid.  
The SMT, with a typical strip pitch of 50--80~$\mu$m, allows a 
precise determination of the primary interaction vertex 
(PV) and an accurate determination of the impact parameter of a 
track relative to the PV~\cite{ip}. The tracker 
design provides efficient charged-particle measurements in the
pseudorapidity region $|\eta| < 3$~\cite{eta_definition}.
The calorimeter consists of a barrel section
covering $|\eta|<1.1$, and two end caps extending to
$|\eta|\approx4.2$.  The muon spectrometer encapsulates the
calorimeter up to $|\eta| = 2.0$ and consists of three layers of drift chambers and two 
or three layers of scintillators~\cite{muon_detector}. A 1.8~T iron toroidal
magnet is located outside the innermost layer of the muon detector.

We select data samples in the electron and muon channels by
requiring an electron with transverse momentum $p_T>20$~GeV
and $|\eta|<1.1$, or a muon with $p_T>20$~GeV and
$|\eta|<2.0$. The leptons are required to be isolated from other 
particles using calorimeter and tracking information. More
details on the lepton identification and trigger requirements are
given in Ref.~\cite{ljetstopo_paper}. $W$ boson candidate events are then 
selected in both channels by requiring missing transverse 
energy, $\met$, in excess of $20$~GeV due to the neutrino. To remove 
multijet background, 
$\met$ is required to be non-collinear with the lepton direction in
the transverse plane. Jets are defined using a cone algorithm~\cite{jet} 
with radius $\Delta{\cal R}=0.5$~\cite{dr_definition}. These events must be
accompanied by four or more jets with $p_T>15$~GeV and rapidity
$|y|<2.5$. After all the above selection requirements are applied, 
we have a total of 231 (277) events in the muon (electron) 
channel. 

We use a secondary vertex tagging (SVT) algorithm to reconstruct displaced
vertices produced by the decay of $B$ hadrons. Secondary 
vertices are reconstructed from two or more tracks satisfying: $p_T>1$~GeV, 
$\geq 1$ hits in the SMT layers, and impact parameter significance
$d_{ca}/\sigma_{d_{ca}}>3.5$. 
A jet is considered as SVT-tagged if it contains a secondary vertex with a
decay length significance $L_{xy}/\sigma_{L_{xy}}>7$~\cite{dl}. The 
determination of the sample composition relies on $b$-tagging,
$c$-tagging, and light flavour tagging efficiencies and uses the
method described in Ref.~\cite{xsec_publ_230}. In order to increase the 
purity of the sample we select only events with two or more SVT-tagged jets. 
In the selected sample of 21 events with two SVT-tagged jets, the largest 
(second largest) background is
$Wb\bar{b}$ (single top quark~\cite{single_top_xs}) production with a 
contribution of $\approx$~5\% ($\approx$~1\%) to the number of selected 
events.

The top or anti-top quark whose $W$ boson decays leptonically (hadronically) 
is refered to as the leptonic (hadronic) top and the associated $b$-quark 
is denoted $b_{\ell}$ ($b_h$). To compute the top quark charge we need to 
{\it i)} decide which of the two SVT-tagged jets are $b_{\ell}$ and $b_h$ 
and {\it ii)} determine if $b_{\ell}$ and $b_{h}$ are $b$- or 
$\bar{b}$-quarks. 
The detected final state partons in the $\ttbar$ candidate events 
comprise the $b_{\ell}$ and $b_h$ quarks, two quarks from the 
hadronically decaying $W$ boson, and one muon or one electron. The four 
highest-$p_T$ jets can be assigned to the set of final state quarks according to
many permutations and there are at least two ways to assign the SVT-tagged 
jets to $b_{\ell}$ and $b_h$.
For each permutation, the measured 
four-vectors of the jets and lepton are fitted to the $\ttbar$ event 
hypothesis, taking into account the experimental 
resolutions and constraining the mass of two $W$ bosons to its measured 
value and the top quark mass to $175$~GeV. We decide which 
of the SVT-tagged jets are 
$b_{\ell}$ and $b_h$ by selecting the permutation with the 
highest probability of arising from a $\ttbar$ event. Studies on simulated 
\ttbar~ show that this gives the correct assignment in about 84\% of 
the events.

We measure the absolute value of the top quark charge on each side of 
the event, given by $Q_1 = | q_{\ell} + q_{b_{\ell}}|$ on the leptonic 
side and $Q_2 = |-q_{\ell} + q_{b_{h}}|$ on the hadronic side. The charge 
of the lepton is indicated by $q_{\ell}$, and $q_{b_{\ell}}$ and 
$q_{b_{h}}$ are the charges of the SVT-tagged jets on the leptonic and 
hadronic side of the event. The charges $q_{b_{\ell}}$ and $q_{b_{h}}$ are 
determined by combining the $p_T$ and charge of the tracks contained within 
a cone of 
$\Delta {\cal R}$=0.5 around the SVT-tagged jet axis. Based on an 
optimization using simulated $\ttbar$ events generated with 
{\sc alpgen}~\cite{ref_alpgen} and 
{\sc geant}~\cite{ref_geant} for a full D0 detector simulation, we define 
an estimator for jet charge 
\(  q_{\text{jet}} = \left( \sum_{i} q_i \cdot p_{T_i}^{0.6} \right) / \left( \sum_{i}p_{T_i}^{0.6} \right) \)
where the subscript $i$ runs over all tracks with $p_T > 0.5$~GeV and 
within 0.1~cm of the PV in the direction parallel to the beam axis.


To determine the expected distributions for the
top quark charges $Q_1$ and $Q_2$, it is crucial to determine 
the expected distributions for $q_{\text{jet}}$ in the case of a 
$b$-quark or a $\bar{b}$-quark jet. In $\approx$5\% of the $\ttbar$ events,
one of the SVT-tagged jets is actually a $c$-quark jet arising from 
$W\to c\bar{s}$ (or its charge conjugate). Therefore we also need 
to determine the expected distribution for $q_{\text{jet}}$ in the case 
of $c$- and $\bar{c}$-quark jets.

We derive the expected distributions of jet charge from dijet 
collider data, enhanced in heavy flavor ($b$ and $c$). 
We select events with exactly two jets, both 
SVT-tagged, with $p_T >$ 15~GeV and $|y|<$2.5. The method requires 
that the two jets are of charge conjugate flavors. To ensure 
this, we enhance $b\bar{b}$ and $c\bar{c}$ produced by flavor 
creation~\cite{rick_field,cdf_b_production_runI,dzero_thesis}, by 
requiring the azimuthal distance between the jets to be larger 
than 3.0 and one jet (designated as $j_1$) to contain a 
muon with $p_T>$ 4~GeV. We refer to this 
sample as the ``tight dijet sample,'' to $j_1$ as the ``tag jet'' 
and to the second jet $j_2$ as the ``probe jet.'' 

The fraction of $c\bar{c}$ events in the tight dijet sample is estimated 
using the distribution of the muon transverse momentum with 
respect to the tag jet axis ($\ptrel$). We fit the $\ptrel$ distribution 
with a sum of two $\ptrel$ templates, one for $b$-quark jets (including both 
prompt and cascade decays) and one for semi-muonic decays inside 
$c$-quark jets. This leads to a fraction $x_c$ of $c\bar{c}$ events of $1^{+2}_{-1}$\%
in the tight dijet sample and since the light flavor tagging efficiency is 
$\approx$15 times lower, we also conclude that the fraction of
lighter flavor jets is negligible. The muon inside the tag jet comes 
either {i)} from a direct $B$ meson decay, {ii)} a 
$B\rightarrow D$ meson cascade decay, {iii)} an oscillated neutral $B$ meson, or {iv)} 
a direct $D$ meson decay. We find that further contribution from indirect 
$D$ meson decay can be neglected. Charge flipping processes {ii)} 
and {iii)} lead to a muon of opposite charge to that of the quark 
initiating the tag jet and therefore of same sign as the quark initiating 
the probe jet. We find, with {\sc pythia}~\cite{ref_pythia} simulated events and 
{\sc evtgen}~\cite{ref_evtgen} for heavy flavor decays, that charge flipping processes 
are $x=\left ( 30 \pm 1 \right )$\% of the $b\bar{b}$ events in the tight dijet sample. 
This fraction is experimentally confirmed by studying charge correlation 
between muons in back-to-back muon-tagged dijet events.

We denote the charge distributions for the probe jet when the muon 
on the tag side is positive or negative as $P_{\mu^+}$ and $P_{\mu^-}$. 
Similarly we define $P_{f}$ to be the charge distribution when the jet 
is of flavor $f=b, \bar{b}, c, \bar{c}$. Given the fractions of 
$c\bar{c}$ events and of charge flipping processes we can write 
\begin{eqnarray}
  P_{\mu^+} & = & 0.69 P_b + 0.30 P_{\bar{b}} + 0.01 P_{\bar{c}} \nonumber \\
  P_{\mu^-} & = & 0.30 P_b + 0.69 P_{\bar{b}} + 0.01 P_c.
\label{eq:tripletag1}
\end{eqnarray}
$P_{\mu^+}$ and $P_{\mu^-}$ are distributions observed in data and are 
admixtures of the quark charge distributions.
Equations~\ref{eq:tripletag1} are not sufficient to extract the four
probability density functions (p.d.f.'s) $P_{f}$. Therefore 
we define a ``loose dijet sample,'' where $j_1$ is not required 
to be SVT-tagged. Using the same techniques as for the 
tight dijet sample, we find that $x_c=\left ( 19 \pm 2 \right )$\% and the 
same fraction of 
charge flipping processes as for the tight dijet sample. We refer to 
$P'_{\mu^+}$ ($P'_{\mu^-}$) 
as the observed p.d.f.'s for $q_{\text{jet}}$ on the probe jet in the loose dijet 
sample, when the tag muon is positive (negative). Thus we can write
\begin{eqnarray}
  P'_{\mu^+} & = & 0.567 P_b + 0.243 P_{\bar{b}} + 0.19 P_{\bar{c}} 
  \nonumber \\
  P'_{\mu^-} & = & 0.243 P_b + 0.567 P_{\bar{b}} + 0.19 P_c. 
\label{eq:musvttag}
\end{eqnarray}
We solve Eqs. 1 and 2 to obtain the $P_{f}$ for \mbox{$b$-,} $\bar{b}$-, $c$-, 
and $\bar{c}$-quark jets.

The $P_{f}$'s are dependent on the jet $p_T$,
since $p_T$ correlates with track multiplicity in the jet, and on the jet
$y$, since the tracking efficiency is rapidity-dependent. Therefore we must
account for the different jet $p_T$ and $y$ spectra between the 
probe jets of the dijet samples and the $b$-quark jets in preselected $\ttbar$ 
events. The $P_{f}$'s obtained above are corrected by weighting the data 
events to the $p_T$ and $y$ spectra of SVT-tagged jets in $\ttbar$ events. 
Figure~1(a) shows the resulting $P_{b}$ and $P_{\bar{b}}$.

We derive the expected distributions for $Q_1$ and $Q_2$ by applying the 
assignment procedure 
between the SVT-tagged jets and the ${b_h}$, ${b_{\ell}}$ 
quarks on simulated $\ttbar$ events using our calculated $P_f$'s. The true 
flavor $f$ of the SVT-tagged jets is determined from the simulation 
information. The values of $q_{b_h}$  and $q_{b_{\ell}}$ are obtained by 
randomly sampling the distribution of $P_{f}$ for the corresponding
flavors. About 1\% of $\ttbar$ candidate events contain a SVT-tagged light-flavor 
jet. In this case the p.d.f. for $q_{\text{jet}}$ is taken from simulation. 
In the case of a $|q|=4e/3$ exotic quark, the expected 
distributions of exotic quark charge are derived by computing 
$Q_1 = | - q_{\ell} + q_{b_{\ell}}|$ and $Q_2 = | q_{\ell} + q_{b_{h}}|$, 
following the same procedure as for the SM top quark. The uncertainty on the 
mass of the top quark~\cite{world_average_top_mass} is propagated as 
a systematic uncertainty.

The expected distributions of $Q_1$ and $Q_2$ for the background 
are obtained by {\it i)} performing the assignment procedure between SVT-tagged 
jets and the $b_h$, $b_{\ell}$ quarks on $Wb\bar{b}$ simulated events, {\it ii)} 
using the true jet flavors $f$ to sample the corresponding $P_{f}$'s. The 
resulting distributions of $Q_1$ and $Q_2$ for the background are added to 
the top charge distributions in the SM and exotic cases. We denote 
$P_{\rm SM}$ ($P_{\rm ex}$) the p.d.f.'s for $Q_1$ and $Q_2$ including 
the background contributions in the SM (exotic) case.

\begin{figure*}[t]
  \setlength{\abovecaptionskip}{5pt}
  \centering
  \includegraphics[width=0.32\textwidth]{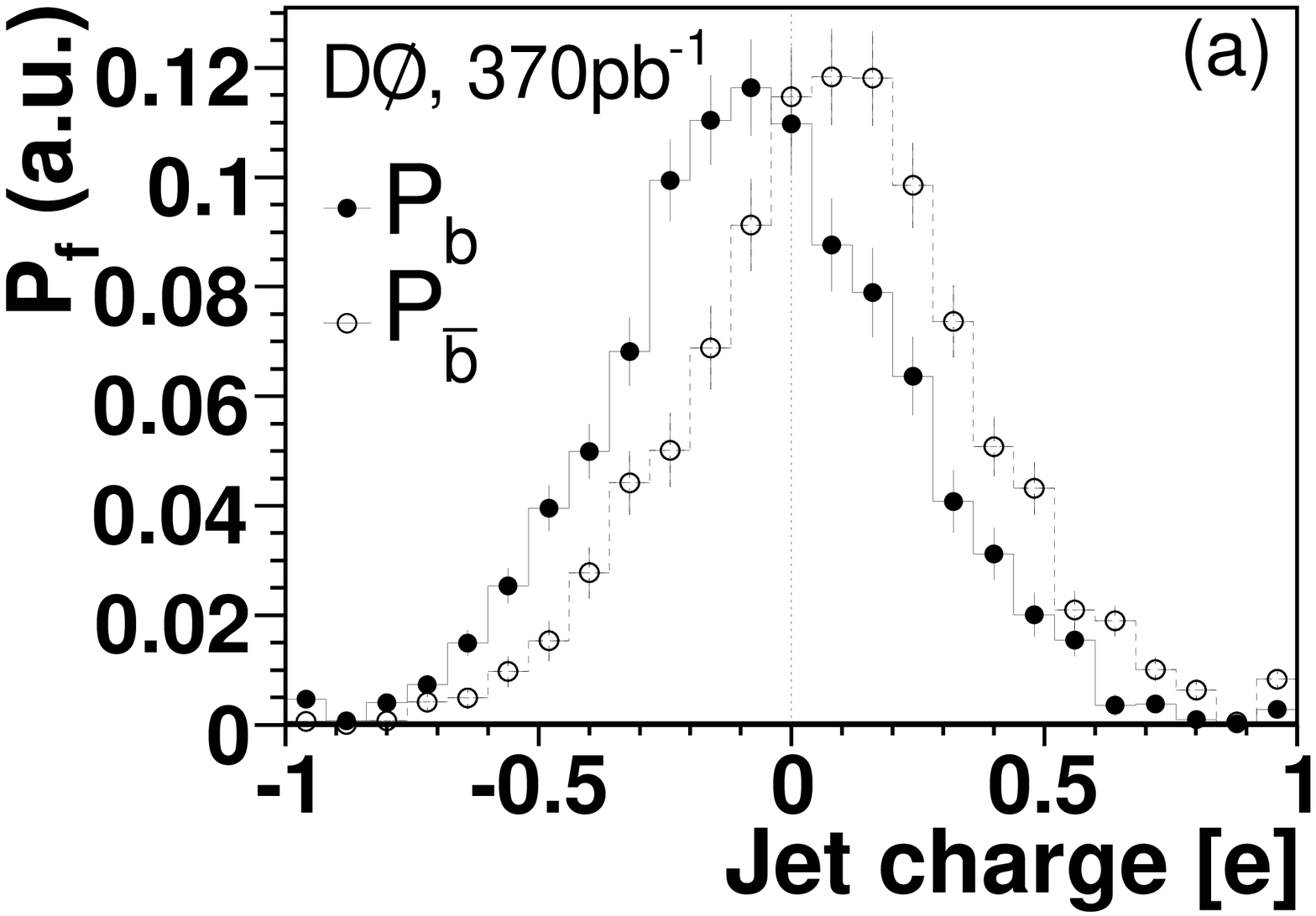}
  \includegraphics[width=0.32\textwidth]{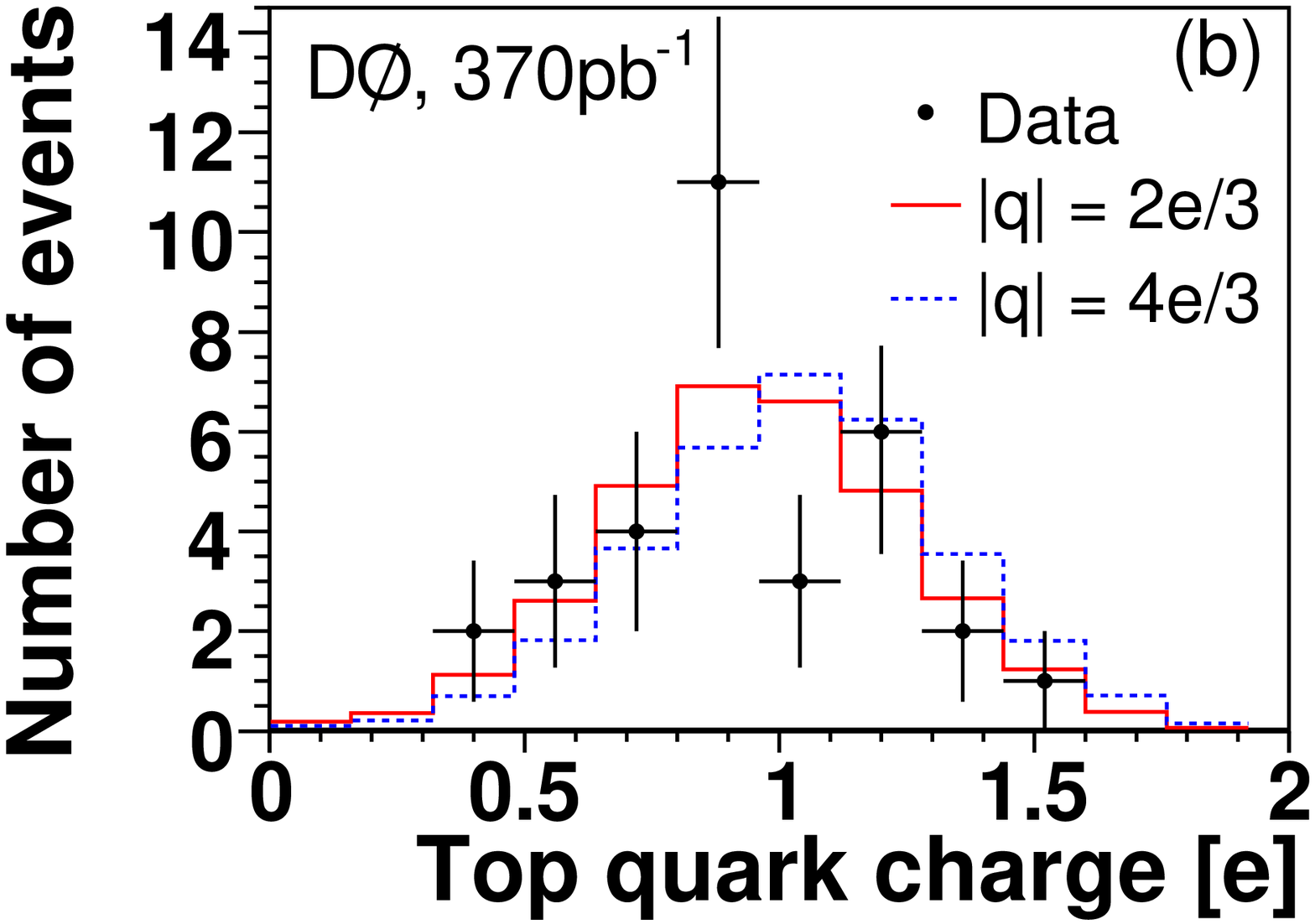}
  \includegraphics[width=0.32\textwidth]{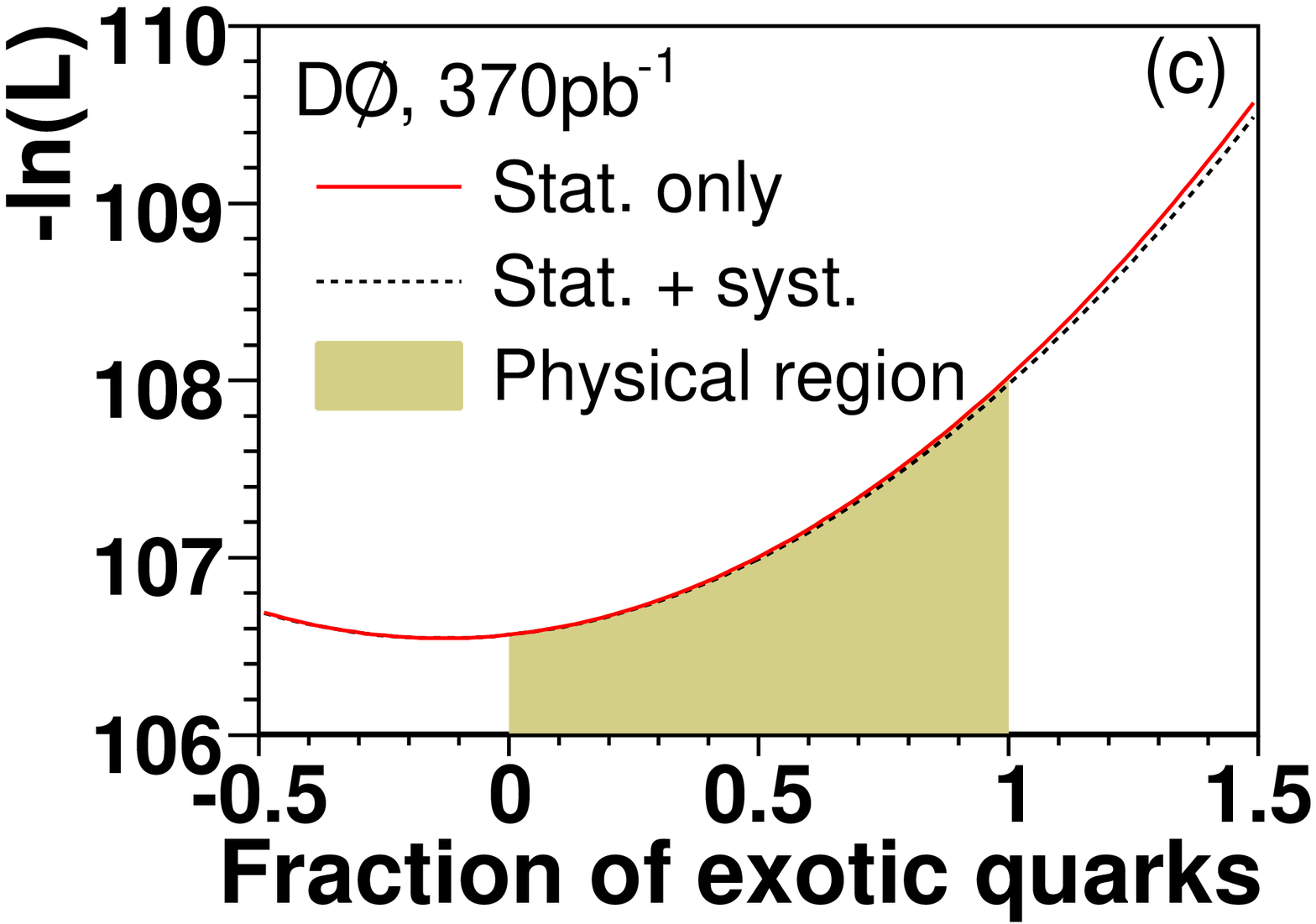}
  \caption{(a) $b$ and $\bar{b}$ jet charge distributions 
    derived from dijet data, (b) the 32 measured values of the top quark 
    charge compared to the expected distributions in the SM and exotic cases, 
    and (c) likelihood fit of the fraction of exotic quark pairs in the selected 
    data sample.}
  \label{fig:topchargedata}
\end{figure*}

For 16 of the 21 selected lepton-plus-jet events, the kinematic fit converges and we can 
assign the SVT-tagged jets to the $b_{\ell}$ and $b_h$ quarks, thus
providing 32 measurements of the top quark charge.
Figure~\ref{fig:topchargedata}(b) shows the 32 observed values of $Q_1$ and 
$Q_2$ overlaid with the SM and exotic charge distributions.

To discriminate between the SM and the exotic hypotheses, 
we form the ratio of the likelihood of the observed set of charges 
$q_i$ arising from a SM top quark to the likelihood for the set of 
$q_i$ arising from the exotic scenario,
$\Lambda = \left[ \prod_i P_{\text{SM}}(q_i)\right] /
\left[ \prod_i P_{\text {ex}}(q_i) \right]$.
The subscript $i$ runs over all 32 available measurements. 
The value of the ratio is determined in data and compared with the
expected distributions for $\Lambda$ in the SM and exotic scenarios.
We find that the observed set of charges agrees well with those of a SM 
top quark. The probability of our observation is 7.8\% in the case 
where the selected sample contains only exotic 
quarks with charge $|q|=4e/3$, including systematic uncertainties. 
Thus, we exclude at the 92.2\% C.L. that the selected data
set is solely composed of an exotic quark with $|q|=4e/3$. 
The corresponding expected C.L. is 91.2\%.
Table~\ref{tab:sysbreakdown} summarizes the dominant systematic 
uncertainties and their cumulative effect on the C.L. 


It is not excluded that the data contain a mixture of two heavy quarks, 
one with $|q|=2e/3$ and one with $|q|=4e/3$. We perform an unbinned maximum 
likelihood fit to the observed set of $q_i$ in data to determine the fraction $\fq$ 
of exotic quark pairs. 
The likelihood of the observed set of $q_i$ can be expressed as 
a function of $\fq$ by
\begin{equation}
  \label{eq:LH}
  L \left(\fq,q\right) = \prod_{i=1}^{N_{\text{data}}} (1-\fq) P_{\rm SM}(q_i) + \fq P_{\rm ex}\left(q_i\right)
\end{equation}
Figure 1(c) shows $-\ln L$ as function of $\fq$. We fit 
$\fq=-0.13\pm0.66\rm(stat)\pm0.11(syst)$, consistent 
with the SM. Using a Bayesian prior equal to one in the 
physically allowed region $0 \leq \fq \leq 1$ and zero otherwise,
we obtain  $0 \leq \fq < 0.52$ at the 68\% C.L. and $0 \leq \fq < 0.80$ 
at the 90\% C.L.

\begin{table}
\begin{tabular}{lcc}
\hline
\hline
Systematic & Observed & Expected \\ 
\hline
Statistical uncertainty only                                 & $95.8$ & $95.3$ \\ 
+ Fraction of $c\bar{c}$ events                              & $95.8$ & $95.2$ \\ 
+ Charge-flipping processes                                  & $95.7$ & $95.2$ \\ 
+ Weighting w.r.t. $p_T$ and $y$ spectra                     & $94.4$ & $94.1$ \\ 
+ Fraction of flavor creation                                & $93.7$ & $93.4$ \\ 
+ Statistical error on $P_f$                                 & $93.3$ & $93.1$ \\ 
+ Jet energy calibration\footnote{Reference~\cite{jes_nim}.} & $92.4$ & $91.8$ \\ 
+ Top quark mass                                             & $92.2$ & $91.2$ \\ 
\hline
\hline
\end{tabular}
\caption{Expected and observed confidence levels as function of the cumulated
  systematic uncertainties.} 
\label{tab:sysbreakdown}

\end{table}

In summary, we present the first experimental discrimination between 
the $2e/3$ and $4e/3$ top quark electric charge scenarios. The observed 
top quark charge is consistent with 
the SM prediction. The hypothesis that only an exotic quark with charge 
$|q|=4e/3$ is produced has been excluded at the 92\% C.L. We also place 
an upper limit of 0.80 at the 90\% C.L. on the fraction of 
exotic quark pairs in the double tagged lepton-plus-jets sample.

%
We thank the staffs at Fermilab and collaborating institutions, 
and acknowledge support from the 
DOE and NSF (USA);
CEA and CNRS/IN2P3 (France);
FASI, Rosatom and RFBR (Russia);
CAPES, CNPq, FAPERJ, FAPESP and FUNDUNESP (Brazil);
DAE and DST (India);
Colciencias (Colombia);
CONACyT (Mexico);
KRF and KOSEF (Korea);
CONICET and UBACyT (Argentina);
FOM (The Netherlands);
PPARC (United Kingdom);
MSMT (Czech Republic);
CRC Program, CFI, NSERC and WestGrid Project (Canada);
BMBF and DFG (Germany);
SFI (Ireland);
The Swedish Research Council (Sweden);
Research Corporation;
Alexander von Humboldt Foundation;
and the Marie Curie Program.
%

%
%


\begin{thebibliography}{99}
\bibitem{ip}
  Impact parameter is defined as the distance of closest approach
  ($d_{ca}$) of the track to the primary vertex in the plane
  transverse to the beamline. Impact parameter significance is
  defined as $d_{ca}/\sigma_{d_{ca}}$, where $\sigma_{d_{ca}}$ is the
  uncertainty on $d_{ca}$.
\bibitem{eta_definition}
  Rapidity $y$ and pseudorapidity $\eta$ are defined as functions of the polar angle
$\theta$ and parameter $\beta$ as
$y(\theta,\beta)=\frac{1}{2}\log(\frac{1+\beta \cos\theta}{1-\beta \cos\theta})$ and
$\eta(\theta)=y(\theta,1)$, where $\beta$ is the ratio of a particle's momentum to its energy.
\bibitem{dr_definition}
  $\Delta {\cal R}$ is defined as a cone in pseudorapidity- and $\phi$-space,
  $\Delta {\cal R} = \sqrt{(\Delta\eta)^2 + (\Delta\phi)^2}$, where $\phi$ is the azimuthal angle.
\bibitem{dl} Decay length $L_{xy}$ is defined as the distance
    from the primary to the secondary vertex in the plane transverse to the beamline.
    Decay length significance is defined as $L_{xy}/\sigma_{L_{xy}}$, where
    $\sigma_{L_{xy}}$ is the uncertainty on $L_{xy}$.
\end{thebibliography}
\end{document}